\documentstyle[prd,aps,epsf]{revtex}
\begin{document}
\draft
\title{Some minimally-variant map-based rules of motion at any speed}
\author{P. Fraundorf}
\address{Physics \& Astronomy, U. Missouri-StL (63121) \\
Physics, Washington U. (63110), \\
St. Louis, MO, USA}
\date{\today }
\maketitle

\begin{abstract}
We take J. S. Bell's commendation of ``frame-dependent'' perspectives to the limit 
here, and consider motion on a ``map'' of landmarks and clocks fixed with respect to a 
single arbitrary inertial-reference frame.  The metric equation connects a 
{\em traveler}-time with {\em map}-times, yielding simple integrals of constant 
proper-acceleration over space (energy), traveler-time (felt impulse), map-time (momentum), 
and time on the clocks of a chase-plane determined to see Galileo's original equations apply 
at high speed.  Rules follow for applying frame-variant and proper forces in context of 
one frame. Their usefulness in curved spacetimes via the equivalence
principle is maximized by using synchrony-free and/or frame-invariant forms
for length, time, velocity, and acceleration. In context of any single system of 
locally inertial frames, the metric equation thus lets us express electric and 
magnetic effects with a single frame-invariant but velocity-dependent force, 
and to contrast such forces with gravity as well.  From physics/9704018.
\end{abstract}




\section{Introduction}

Motion at any speed can be studied without a need for transforming between
objects in relative motion, using information from the space-time metric
equation itself concerning the rate at which moving clocks slow with respect
to fixed clocks, if simultaneity and distances are considered only with
respect to a single inertial reference (or ``map'') frame. This works
without the need for transforming, since choosing a map-frame unambiguously
defines both distances between objects, and simultaneity between events: two
things which relativity has shown to be frame-dependent.

This approach was not used in early papers \cite{Einstein}, perhaps
because an early inspiration for the ideas was the postulate that light
recognizes no physically-special frame-of-motion. However, when the focus
turns to solving problems, it is a good idea \cite{Bell} to select a
convenient, if otherwise arbitrary, frame of reference for drawing maps and
evaluating variables. For example, when mapping the solar system we might
choose a map-frame at rest with respect to the sun, or the solar system's
center of mass. Likewise, general relativistic ``embedding diagrams'' of 
space-time curvature require a selected frame of motion as well \cite{Thorne}. 
Merely specifying a momentum or a kinetic energy requires the same.

Once this choice is made, the metric equation allows us to draw out 
predictions for any motions that we wish, and to express the results
concretely in terms of map-position, and time-elapsed on the traveler and
map clocks. Moreover, by choosing the motion of interest to involve constant
acceleration ``felt'' by the traveling object of interest, we find that
Newton's second law at high speed breaks into two integrals, one for
time-elapsed on traveler clocks, and a second one (related to momentum
conservation) for time-elapsed on the map. A work-energy relation very
similar to the classical one also emerges, as does a way to put Galileo's
geometrically-derived equations for constant acceleration to use at any speed.  

This conceptual approach is of special interest for engineering physics.  
For example workers taught to specify the clock whenever 
discussing elapsed times can then confidently deal with motions at any speed,  
concretely in context of any user-specified reference frame 
\cite{anticipation}, and in that context a velocity-dependent and frame-invariant 
force law predicts electric and magnetic forces as one.  The contrasting origins 
of gravity in curvature-related proper-acceleration may be easier 
to see as well.

\section{The basics}

First, a word about notation. Many of the relationships we discuss here are
easy to see in 4-vector form. In fact, a set of 4-vector equations for
motion under constant proper acceleration has been consulted often in the
writing of this paper. However the objective is to make calculations
involving motion at any speed as concrete as possible. Our strategy
is based deeply in the metric, but is explicitly frame-dependent rather 
than coordinate-free.  Although space-time symmetries are somewhat hidden 
as a result, the approach is useful because our {\em view of the world} 
is inherently frame-specific.  Thus we experience time differently than we do 
space, clocks and meter-sticks are constructed differently, and the equations used 
by scientists and engineers solving practical relativistic problems will likely 
distinguish between units for distance and time.  Hence the spirit of what follows.

\subsection{The kinematical and dynamical postulates}

Because the focus is on motion at any speed, but not on relative motion, we
begin with the metric equation rather than with Lorentz transforms or
postulates about light. A teacher might, for example, begin by saying: ``To
understand how magnetism results from electric currents in neutral wires,
and how lightspeed avoids dependence on the earth's direction of travel,
Einstein postulated \cite{Einstein} (in terms introduced by Minkowski 
along the road toward general relativity \cite{Einstein2}) that 
the time $d\tau $ elapsed on the clocks of a traveler relates to the 
time and position changes ($dt$, $d{\bf x}$) seen by map-based observers 
via the Pythagoras-like metric equation:

\begin{equation}
(cd\tau )^2=(cdt)^2-(dx)^2-(dy)^2-(dz)^2\text{,}  \label{metric}
\end{equation}
where $c$ is the speed of light.'' We say here that this equation
constitutes the ``kinematical assumption'' in flat space-time.

For each of the two times in this equation, we define two velocities or
time-derivatives of {\em position on the map}. The derivative with respect
to {\em map-time} is the usual {\em coordinate-velocity} ${\bf v}\equiv 
\frac{d{\bf x}}{dt}$, while the derivative with respect to {\em traveler-time%
} is called {\em proper-velocity} ${\bf w}\equiv \frac{d{\bf x}}{d\tau }$ [%
\cite{SearsBrehme,Shurcliff}]. It is easy from the metric equation to show
that the ${\em rate}$ at which map-time changes per unit traveler-time, or
in other words the ``{\em speed-of-map-time}'' $\gamma $ if we define
simultaneity from the perspective of the map-frame, is simply:

\begin{equation}
\gamma \equiv \frac{dt}{d\tau }=\frac 1{\sqrt{1-\left( \frac vc\right) ^2}}=%
\sqrt{1+\left( \frac wc\right) ^2}\text{.}  \label{gamma}
\end{equation}
The above variable definitions take steps toward making application of the
integrals of accelerated motion, to be discussed in the next section, as
simple and concrete as possible.

Also relevant to our choice of variables is the ``dynamical assumption''.
One might describe this simply by saying: ``Einstein further noted that an
object of mass $m$ may then have a rest-energy $E_o\equiv mc^2$, related to
total energy $E$ (kinetic plus rest) and momentum $p$ by the relation

\begin{equation}
E^2=E_o^2+\left( pc\right) ^2\text{.''}  \label{energy_momentum}
\end{equation}
This relation follows by multiplying the metric equation above by $(mc)^2$
and dividing by $d\tau $ twice, if one postulates that the quantities momentum $%
p\equiv mw$ and total energy $E\equiv \gamma mc^2$ are conserved in all
isolated interactions. This dynamical assumption has of course been shown to
work very well in practice.

\subsection{Variables to monitor changes in velocity}

For describing motion at any speed, we introduce three other classes of
variables: a pair of velocity functions, a pair of accelerations, and the
``Galilean-kinematic'' velocity/time pair. The first velocity function is
the {\em parallel-rapidity} (or hyperbolic velocity-angle) $\eta _{\parallel
}\equiv \sinh ^{-1}[\frac{w_{\parallel }}c]$. This operates on $w_{\parallel
}$, the component of proper-velocity {\em in} the direction of
coordinate-acceleration. The second velocity-component function might be
called ``{\em gamma-perp}'', and is defined as $\gamma _{\perp }\equiv \frac 
1{\sqrt{1-\left( v_{\perp }/c\right) ^2}}$. This operates on that component
of coordinate-velocity, namely $v_{\perp }$ or ``v-perp'', which is {\em %
orthogonal to} the direction of coordinate-acceleration. This quantity is
that part of the speed-of-map-time $\gamma \equiv \frac{dt}{d\tau }$ that 
{\em would result} from only the component of coordinate-velocity
perpendicular to the direction of velocity change. Because $\gamma _{\perp }$
depends entirely on the velocity-component perpendicular to the direction of
velocity-change, it will be constant iff the {\em direction} of acceleration
is constant. Also note that when the motion is uni-directional (e.g. in 1+1D
special relativity), $v_{\perp }=0$, $\gamma _{\perp }=1$, and we can define
rapidity $\eta \equiv \sinh ^{-1}[\frac wc]$.

Concerning accelerations, {\em coordinate-acceleration} ${\bf a}\equiv \frac{%
d{\bf v}}{dt}=\frac{d^2{\bf x}}{dt^2}$ is the only acceleration discussed in
classical descriptions of motion \cite{Newton,Chandra}. Given two times to
keep track of, one might also consider proper, and mixed, time derivatives
of coordinate position. As we show later, however, of most use is the
frame-invariant acceleration ``felt'' by the traveler, i.e. the {\em %
proper-acceleration} \cite{TaylorWheeler}. This acceleration is not the
second time-derivative of position with respect to either map or traveler
time! Rather, it's magnitude is the rate at which parallel rapidity changes
per unit traveler-time ($\alpha \equiv \frac{d\eta _{\parallel }}{d\tau }$),
while its vector relation to coordinate-acceleration is: ${\bf \alpha }=%
\frac{\gamma ^3}{\gamma _{\perp }}{\bf a}$ \cite{twoclock}. This
acceleration enters directly into integrals of the motion involving the
conserved quantities energy and momentum, as we will see, and hence is the
acceleration most deeply connected to the dynamics of motion.

\subsection{Galileo's ``chase-plane''}

The last set of variables are the ``Galilean-kinematic'' time-velocity pair.
These are based on the behavior of physical clocks in a chase-plane, whose
velocity is linked to that of our traveler so that proper-acceleration is
the second derivative of traveler-position on the map, with respect to
chase-plane time. If we use $T$ to denote time on the clocks of such a chase
plane, and $V\equiv \frac{dx_{\parallel }}{dT}$ to denote the rate of change
of map-position for our traveler per unit time on chase-plane clocks, then
we require that $\alpha =\frac{dV}{dT}$.

This ``Galilean-kinematic'' velocity $V$, like proper-velocity $w$, involves
the rate of change of a traveler's position, with respect to landmarks and
meter-sticks at rest in a map-frame, per unit time on clocks not stationary
in that map frame. These ``non-coordinate'' velocities are examples of the
infinite set of such physical non-coordinate velocities which may be used to
monitor the motion of objects on reference-frame maps. The requirement
above, for proper-acceleration as a second-derivative of map-position, is
met for the clocks in a chase-plane traveling with total velocity $v^{\prime
}\equiv \frac{dx_{cp}}{dt}=c\sqrt{1-\frac{\left( v/c\right) ^2}{2(\gamma
/\gamma _{\perp }-1)}}$, and with perpendicular component matching the
traveler (i.e. $v_{\perp }^{\prime }=v_{\perp }$) \cite{Noncoord}. Here
primed quantities refer to chase-plane motion with respect to the map-frame,
while unprimed quantities refer to traveler motion in the map-frame, as
described above.

One can show that the motion of this chase-plane is intermediate to that of
traveler and map, so that in the classical limit when differences in
traveler and map time are negligible, these chase-plane equations provide an
accurate and simple approximation: namely the approximation discovered by
Galileo well before Newton's birth \cite{Galileo}. Remarkably, they also
provide an accurate description of motion {\em at any speed} on a
reference-frame map, monitored with clocks in the chase-plane defined here!

\subsection{Inter-velocity conversions}

It may be helpful to summarize the foregoing kinematic relations between
parameters related to velocity with a single string of equalities. If we use 
$\gamma $ $\equiv \frac{dt}{d\tau }=\frac E{mc^2}$ as the dependent quantity
(since it is the one ``velocity parameter'' which discards sign
information), we can add terms for parallel-rapidity and Galilean-velocity
to the equality string in equation \ref{gamma}: 
\begin{equation}
\gamma =\gamma _{\perp }\cosh \left[ \eta _{\parallel }\right] =\gamma
_{\perp }\left( 1+%
{\textstyle {1 \over 2}}
\left( 
{\textstyle {V \over c}}
\right) ^2\right) \text{.}  \label{g3string}
\end{equation}
For the special case of unidirectional motion, the $\gamma $ string
simplifies to 
\begin{equation}
\gamma =\cosh \left[ \eta \right] =1+%
{\textstyle {1 \over 2}}
\left( 
{\textstyle {V \over c}}
\right) ^2\text{.}  \label{g1string}
\end{equation}
Except for the above-mentioned information on velocity sign, these strings
contain all of the information on velocity inter-conversions needed for the
solution of constant acceleration problems at any speed. Conversions in
closed form, which preserve the sign-information, follow directly as well.

\section{The integrals of constant acceleration.}

As H. Minkowski suggests \cite{Minkowski}, the metric equation alone is
adequate to derive the integrals of motion under constant accelerations of
any type, just as integrals of motion in the classical limit follow from the
kinematic connection between acceleration, velocity, position, and time
there as well \cite{Schutz}. The dynamics, which of course tie the metric
equation into applications, concern not these integrals themselves, but
rather the way that some variables are conserved when objects interact. For
example, in classical kinematics, the time (or momentum) integral, namely $%
\Delta v=a\Delta t$, and the distance (or work-energy) integral, namely $%
\Delta \left( \frac 12v^2\right) =a\Delta x$, both follow from our
definitions of velocity, acceleration, and the $c\rightarrow \infty $ metric
association between time and distance alone. The dynamical usefulness of the
equations, however, only becomes clear when we associate $mv$ and $\frac 12%
mv^2$ with the conserved quantities momentum, and energy, respectively.

Similarly, in the case of flat space-time, one can derive integrals of any
type of acceleration using kinematical assumptions (i.e. the metric
equation) alone. Because of our occupation with quantities conserved in
dynamical interaction, integrals of constant coordinate-acceleration are of
little interest \cite{French}. Integrals of constant frame-invariant
``felt'' (i.e. proper) acceleration are, therefore, the focus here.

\subsection{The integrals of constant proper acceleration in (3+1)D.}

Before considering integrals of the motion for constant proper-acceleration $%
{\bf \alpha }$, let's review the classical integrals of motion for constant
coordinate-acceleration ${\bf a}$. These can be written as $a\simeq \frac{%
\Delta v_{\parallel }}{\Delta t}\simeq \frac{\Delta (\frac 12v^2)}{\Delta
x_{\parallel }}$. The first of these is associated with conservation of
momentum in the absence of acceleration, and the second with the work-energy
theorem. These may look more familiar in the form $v_{\parallel f}\simeq
v_{\parallel i}+a\Delta t$, and $v_{f}^2\simeq v_{i}^2+2a\Delta x_{\parallel }$.

Given the variables introduced above, four simple integrals of the proper
acceleration can be obtained by a procedure which works for integrating
other non-coordinate velocity/time expressions as well \cite{Noncoord}.
The resulting integrals are summarized in compact form, like the classical
ones in the pargraph above, as

\begin{equation}
\alpha =\gamma _{\perp }\frac{\Delta w_{\parallel }}{\Delta t}=c\frac{\Delta
\eta _{\parallel }}{\Delta \tau }=\frac{c^2}{\gamma _{\perp }}\frac{\Delta
\gamma }{\Delta x_{\parallel }}=\frac{\Delta V}{\Delta T}\text{.}
\label{integrals}
\end{equation}
Note that both $v_{\perp }$ and the ``transverse time-speed'' $\gamma
_{\perp }$ are constants. Hence proper-velocity and longitudinal momentum $%
p_{\parallel }\equiv mw_{\parallel }$ change uniformly with map-time,
parallel rapidity (and impulse delivered from the traveler's view) change
uniformly with traveler-time, energy $E\equiv \gamma mc^2$ changes at a
uniform rate with increase in map-distance, and Galilean-kinematic velocity
changes uniformly with increase in chase-plane time.

Using these equations, one can solve constant acceleration problems at any
speed, and can examine the quantitative behavior of any of the velocities
and accelerations, as well as their relationship to dynamical quantities
like momentum and energy. We draw out some of these relationships in the
last section of this paper, where we attempt a practical summary which at
the same time mirrors underlying relationships.

\subsection{The integrals of constant proper acceleration in (1+1)D.}

If motion is only in the direction of acceleration, $\gamma _{\perp }$ is 1,
and the above integrals simplify to:

\begin{equation}
\alpha =\frac{\Delta w}{\Delta t}=c\frac{\Delta \eta }{\Delta \tau }=c^2%
\frac{\Delta \gamma }{\Delta x}=\frac{\Delta V}{\Delta T}\text{.}
\label{1ntegrals}
\end{equation}
This string of equalities, in combination with the velocity
inter-conversions in the previous section, provide equations for solving
uni-directional constant acceleration problems at any speed.

One might consider 11 variables at the heart of {\em any} uni-directional
problem to be the frame-invariant acceleration $\alpha $, the map-distance
traveled $\Delta x$, elapsed-times in context of map, traveler, and/or
chase-plane clocks ($\Delta t$, $\Delta \tau $, $\Delta T$), and initial and
final velocities expressed inter-changably as coordinate-velocities ($v_i$, $%
v_f$), proper-velocities ($w_i$, $w_f$), or Galilean-kinematic velocities ($%
V_i$, $V_f$). Only three of these variables can be specified independently.

Because of the interchangability of velocity types, this leaves us with 25
types of independent variable assignments, 3 of which involve no
elapsed-times given, 12 of which involve only a single given elapsed-time, 9
of which involve two given elapsed-times, and the last of which involves all
three elapsed-times as input. Although none of these problems is difficult
to solve numerically, we have yet to identify closed-form solutions to one
of the 12 single-time problem types, and 9 of the 10 which involve more than
one input time-elapsed.

\section{Some ``map-based'' rules of motion at any speed.}

In what follows, we summarize the consequences of these observations in a
form designed to highlight their physical meaning for motion in flat
space-time. The equivalence principle for curved space-times, however, 
as well as the frame-invariance of many of the quantities used, extends 
their range of applicability considerably.  The first three items in the 
resulting list of ``rules'' shows similarities with Newton's three laws of 
motion \cite{Newton}, reflecting elements of Newton's world that survive 
even at high speeds.
Unlike Newton's laws, the items here are not all in the form of
postulates. Rather, the kinematic postulate is introduced in I (before rates
of change of velocity are considered), and some of its consequences when
considering rates of velocity change are ennumerated in II, and VII. The
postulate of momentum conservation is introduced in III, with some of its
consequences at high speed ennumerated in IV. Lastly, the postulate of
energy conservation (which Maxwell and others said was implicit in Newton's
laws as well \cite{Chandra}) is introduced explicitly in V, with some
consequences drawn out in VI.

The only quantity from above not redefined in any one place below, perhaps
because it is useful in so many places, is ``gamma-perp'' $\gamma _{\perp
}\equiv \frac 1{\sqrt{1-\left( v_{\perp }/c\right) ^2}}$, i.e. the
``speed-of-map-time'' that would result from the component of
coordinate-velocity perpendicular to the direction of velocity change, taken
alone. We should also add that assigning colors according to
frame-invariance (e.g. red: variant at all speeds; green: variant only at
high speeds; blue: invariant) can be an enlightening exercise.

\subsection{Poly-directional motion in flat space-time.}

{\bf I.} {\bf Motion without forces: }In the absence of external forces,
objects follow geodesics in space-time. For the special case of the flat
metric, this means that an object at rest continues at rest, and an object
in motion continues in motion. When one defines time measurements (hence
simultaneity) and {\em all} distances used from the perspective of a single
``map'' frame, this means that such objects (or travelers) have constant 
{\em coordinate-velocity} ${\bf v}\equiv \frac{d{\bf x}}{dt}$, constant {\em %
proper-velocity} ${\bf w}\equiv \frac{d{\bf x}}{d\tau }$, and a constant
rate $\gamma \equiv \frac{dt}{d\tau }=\frac 1{\sqrt{1-\left( v/c\right) ^2}}=%
\sqrt{1+\left( \frac wc\right) ^2}$ at which {\em map-time} $t$ changes per
unit time $\tau $ on object (or traveler) clocks.

{\bf II. The proper-time/impulse integral: }The invariant force ${\bf F}_o$ 
{\em felt} by an accelerated object may be obtained by multiplying that
object's proper-acceleration ${\bf \alpha }$ by its rest-mass $m$, i.e. $%
{\bf F}_o=m{\bf \alpha }$. This {\em proper-acceleration} ${\bf \alpha }$ is
in the direction of the object's coordinate-acceleration ${\bf a}\equiv 
\frac{d{\bf v}}{dt}$. Its invariant magnitude is lightspeed $c$ times the
rate that the {\em parallel-rapidity}, or hyperbolic arcsine of the
unit-free proper-velocity in the direction of that acceleration, changes per
unit time $\tau $ on the clocks of the object, i.e. $\alpha =c\frac{\Delta
\eta _{\parallel }}{\Delta \tau }=\left( \frac{\gamma ^3}{\gamma _{\perp }}%
\right) a$, where $\eta _{\parallel }=\sinh ^{-1}[\frac{w_{\parallel }}c%
]=\tanh ^{-1}[\gamma _{\perp }\frac{v_{\parallel }}c]=\cosh ^{-1}[\frac 
\gamma {\gamma _{\perp }}]$.

{\em Aside:} At rest on the earth, all objects experience acceleration against
the gravitational curvature of space-time, directed toward center of our planet. 
The way this curvature arises from the web of interaction with participating 
masses is in general a matter for the mathematics of general relativity, although 
calculus can go a long way toward drawing out its consequences in simple 
situations \cite{Taylor2}.  Regardless of this, rule {\bf II} (with a bit of 
help from the equivalence principle) says that such  
acceleration gives rise to a proper-force proportional to the mass $m$ of the 
object involved.  The way in which forces combine from any vantage point, and 
hence their dynamical consequences, is best treated from a specific frame 
of reference.  The next 4 rules do this, thus offering and implementing separate 
``map-based'' rules for momentum and energy conservation from any inertial point 
of view.  

As we mentioned above in discussing differences between the classical (low-speed) 
rules and ``map-based'' rules that work at any speed:  
Just as classical time must be separated into map-time and 
traveler-time, so the classical 2nd law \cite{Newton} separates into the proper-time 
integral of rule {\bf II} here, and a map-time integral associated with 
conservation of momentum in rule {\bf IV}. Conservation of energy (whose explicit 
treatment developed more fully {\em after} Newton \cite{Chandra}) is then given 
statement, and integration, in rules {\bf V} and {\bf VI} respectively.

{\bf III. Action-reaction: }The vector momentum ${\bf p}$ of an object is
equal to its rest-mass $m$ times its proper velocity ${\bf w}$. For every
transfer of momentum from one object to another, there is an equal and
opposite transfer of momentum from the other back to the first.

{\bf IV. The map-time/momentum integral: }The rate of momentum transfer from
object $i$ per unit change in map-time $t$ defines a {\em frame-variant force%
} ${\bf F}_i$ of object $i$, on our moving object. The sum of frame-variant
forces on our object then equals the change in proper-velocity ${\bf w}$ per
unit map-time $t$, times rest-mass $m$, or ${\bf F}_{net}=\sum {\bf F}_i=m%
\frac{d{\bf w}}{dt}$. This net frame-variant force ${\bf F}_{net}\parallel d%
{\bf w}$ aligns with the ``felt'' force ${\bf F}_o\parallel d{\bf v}$ when $%
v\ll c$, and with velocities ${\bf v}\parallel {\bf w}$ for $v\approx c$.
More specifically, its component in the acceleration direction is integrable
for constant ``felt force'' ${\bf F}_o$, and simply $\frac{\Delta
p_{\parallel }}{\Delta t}=\frac{{\bf F}_o}{\gamma _{\perp }}$, while its
component orthogonal is $\frac{dp_{\perp }}{dt}=\gamma _{\perp }\frac{%
v_{\perp }}c\frac{v_{\parallel }}c{\bf F}_o$. The angles each from $d{\bf w}$
and from ${\bf v}$, to $d{\bf v}$, approach $\tan ^{-1}[\gamma _{\perp }%
\frac{v_{\perp }}c]$ from opposite directions as $v$ approaches $c$. This is
where the dynamical part of Newton's 2nd law has been moved, following the
time-split between map and traveling-object clocks at high speed.

{\bf V. The total energy, and dispersion, relations: }For every transfer of
energy from one object to another, there is a loss of energy from the first.
According to Chandrasekhar \cite{Chandra}, this may have been implicit in
Newton's action-reaction law. Total energy and momentum are related by $E=%
\sqrt{(mc^2)^2+(pc)^2}=\gamma mc^2$.

{\bf VI. The work or displacement/energy integral: }The rate at which an
object's energy changes, per unit map-distance traveled in the direction of
the ``felt'' force, equals the magnitude of the ``felt'' force $F_o$ times
gamma-perp, i.e. $\frac{\Delta E}{\Delta x_{\parallel }}=\gamma _{\perp }F_o$%
. In fact there are three quite distinct integrals associated with energy
which are good at any speed, namely $\Delta E=\int \gamma _{\perp }{\bf F}%
_o\bullet d{\bf x}=\int {\bf F}\bullet d{\bf x}=\int vdp$. Note that the
angles in the two dot-product expressions are different!

{\bf VII. The Galilean chase-plane integrals: }If one considers time $T$ on
the clocks of a chase-plane following our object with the same transverse
velocity (i.e. $v_{\perp }^{\prime }=v_{\perp }$) and with total velocity of 
$v^{\prime }=c\sqrt{1-\frac{\left( v/c\right) ^2}{2(\gamma /\gamma _{\perp
}-1)}}$, then uniformly accelerated motion {\em at any speed} is described
by Galileo's two simple integrals for motion under constant acceleration,
namely: $\Delta V=\alpha \Delta T$ and $\Delta \left( V^2\right) =2\alpha
\Delta x$ \cite{Noncoord}. Since chase-plane motion parallel to the
acceleration-direction is intermediate to that of object and map-frame, at
speeds low enough to make elapsed-times for object and map similar, these
equations predict the behavior of their clocks, and parallel velocity
components as well.

\subsection{Conversions between proper-force and variant-force.}

If one is given variant-force ${\bf F}$ (and it's direction) rather than
proper-force ${\bf F}_o$, calculating vector components (like v-perp) that
relate to the proper-force direction is less direct. A helpful strategy is
to determine first the proper-velocity component transverse to the
variant-force, namely $w_t$. To minimize confusion, we use the adjectives
``transverse'' ($t$) and ``longitudinal'' ($l$) for orientations with
respect to {\em variant-force}, while ``perpendicular'' ($\perp $) and
``parallel'' ($\parallel $) are used for orientations with respect to {\em %
proper-force}. With help from a ``transverse-gamma'' and a ``gamma-both'',
as well as the ``gamma-perp'' defined already, the following simple
interconversions to and from variant and proper components then obtain:

\[
\gamma _{\perp }\equiv \frac 1{\sqrt{1-\left( \frac{v_{\perp }}c\right) ^2}}=%
\sqrt{\gamma _t^2+\left( \frac{w_t}c\frac{w_l}c\frac 1{\gamma _t}\right) ^2} 
\]

\begin{eqnarray*}
\gamma _t &\equiv &\sqrt{1+\left( \frac{w_t}c\right) ^2}=\frac 1{\sqrt{%
\left( \frac 1{\gamma _{\perp }}\right) ^2+\left( \gamma _{\perp }\frac{%
v_{\perp }}c\frac{v_{\parallel }}c\right) ^2}} \\
\gamma _{both} &\equiv &\gamma _t\gamma _{\perp }=\sqrt{\gamma _t^4+(\frac{%
w_t}c\frac{w_l}c)^2}=\frac 1{\sqrt{\left( \frac 1{\gamma _{\perp }}\right)
^4+\left( \frac{v\perp }c\frac{v_{\parallel }}c\right) ^2}}
\end{eqnarray*}

\[
\gamma \equiv \frac{dt}{d\tau }=\sqrt{1+\left( \frac wc\right) ^2}=\frac 1{%
\sqrt{1-\left( \frac vc\right) ^2}}\text{.} 
\]

The above quantities are for convenience defined {\em both} in terms of
proper-velocity components projected in the variant-force direction, and
coordinate-velocity components projected in the proper-force direction.
Using these quantities, $v$-perp can be calculated from $w$-transverse, and
vice-versa, since:

\begin{eqnarray*}
v_{\perp } &\equiv &\frac{dx_{\perp }}{dt}=w_t\left( \frac \gamma {\gamma
_{both}}\right) \\
w_t &\equiv &\frac{dx_t}{d\tau }=v_{\perp }\left( \frac{\gamma _{both}}\gamma
\right)
\end{eqnarray*}

Lastly, both the magnitude and direction of proper-force ${\bf F}_o$ can be
given in terms of transverse and longitudinal unit vectors $\widehat{{\bf i}%
_t}$ and $\widehat{{\bf i}_l}$ oriented with respect to the variant-force $%
{\bf F}$.

\[
{\bf F}_o=F\gamma _t\widehat{{\bf i}_{\parallel }}\text{, where }\widehat{%
{\bf i}_{\parallel }}=\frac{\gamma _t^2\widehat{{\bf i}_l}+\frac{w_t}c\frac{%
w_l}c\widehat{{\bf i}_t}}{\gamma _{both}}\text{.}
\]

Similarly, the magnitude and direction of the variant-force ${\bf F}$
follows in terms of perpendicular and parallel unit vectors $\widehat{{\bf i}%
_{\perp }}$ and $\widehat{{\bf i}_{\parallel }}$ oriented with respect to
the proper-force ${\bf F}_o$:

\[
{\bf F}=F_o\frac 1{\gamma _t}\widehat{{\bf i}_l}\text{, where }\widehat{{\bf %
i}_l}=\left( \frac 1{\gamma _{\perp }^2}\widehat{{\bf i}_{\parallel }}+\frac{%
v_{\perp }}c\frac{v_{\parallel }}c\widehat{{\bf i}_{\perp }}\right) \gamma
_{both}\text{ .}
\]

\subsection{Forces, like electromagnetism, whose variant-force depends on
position}

A fixed charge $Q$ acts on a moving charge $q$ via the radial frame-variant
force ${\bf F}=k\frac{qQ}{r^2}\widehat{{\bf r}}$. The frame-invariant
proper-force, using the conversions above, is then simply:

\[
{\bf F}_o=k\frac{qQ}{r^2}\sqrt{1+\left( \frac{w_\phi }c\right) ^2}\widehat{%
{\bf i}}_{\parallel }\text{, where }\widehat{{\bf i}}_{\parallel }=\frac{%
\left[ 1+\left( \frac{w_\phi }c\right) ^2\right] \widehat{{\bf r}}+\left[ 
\frac{w_\phi }c\frac{w_r}c\right] \widehat{{\bf \phi }}}{\sqrt{\left[
1+\left( \frac{w_\phi }c\right) ^2\right] ^2+\left[ \frac{w_\phi }c\frac{w_r}%
c\right] ^2}}\text{.} 
\]
This latter force is the proper-force seen by {\em any} observer. Hence one
can convert from this common proper-force to the variant-force from any
other frame of motion, using the equations above. If that frame of motion is
one in which magnetic forces between the charges are expected, then this
proper-force will from that vantage point include the magnetic force
components as well. This single equation thus describes electricity and
magnetism {\em from any point of view}, with the caveat of course that the
independent variables are written here in terms of distances measured in the
rest frame of the charge.

We can be more specific in 4-vector terms by pointing out that the
variant-force is associated with a 4-vector potential which for our charge
has components $\{c\Phi ,A_x,A_y,A_z\}=$ $\{ck\frac Qr,0,0,0\}$. These
quantities are related to electric and magnetic fields by the usual
relations ${\bf E\equiv }-{\bf \nabla }\Phi -\frac 1c\frac{\partial {\bf A}}{%
\partial t}=k\frac Q{r^2}\widehat{{\bf r}}$, and ${\bf B}\equiv {\bf \nabla }%
\times {\bf A}=0$. When observed from a different frame, of course, the
4-vector potential will be rotated so as to have space-like components. Thus
the charge $Q$ will exhibit a magnetic field, even though its variant and
proper-forces are both purely Coulombic in nature.

It is easy to see that these equations allow us to predict a
velocity-dependent proper-force, and spacelike components to the vector
potential when seen from a moving frame, whenever the variant-force is a
function solely of position while the agent of that force is at rest. Viewed
in this way, the flat space-time metric requires {\em analogs to magnetism
for all conservative forces}. With such a strong requirement for ``magnetic
effects'' associated with any conservative force, the formal convenience
in Maxwell's equations of free magnetic charges appears (in the absence of
evidence for such charges) to be a formalism curiosity and no more.

\subsection{Uni-directional motion at any speed.}

Those items in the list above which do not introduce postulates (i.e. II,
IV, VI, and VII) simplify sufficiently, for uni-directional motion, that we
rewrite them here specifically for that case:

{\bf IIu. The uni-directional proper-time/impulse relation: }The force ${\bf %
F}_o$ {\em felt} by an accelerated object may be obtained by multiplying
that object's proper-acceleration ${\bf \alpha }$ by its rest-mass $m$, i.e. 
${\bf F}_o=m{\bf \alpha }$. This {\em proper-acceleration} ${\bf \alpha }$
has a frame-invariant magnitude equal to lightspeed $c$ times the rate that
the {\em rapidity}, or hyperbolic arcsine of the unit-free proper-velocity,
changes per unit time $\tau $ on the clocks of the object, i.e. $\alpha =c%
\frac{\Delta \eta }{\Delta \tau }=\gamma ^3a$, where $\eta =\sinh ^{-1}[%
\frac wc]=\tanh ^{-1}[\frac vc]=\cosh ^{-1}[\gamma ]$.

{\bf IVu. The unidirectional map-time/momentum relation: }The rate of
momentum transfer from object $i$ per unit change in map-time $t$ defines a 
{\em frame-variant force} ${\bf F}_i$ of object $i$, on our moving object.
The sum of frame-variant forces on our object then equals the change in
proper-velocity ${\bf w}$ per unit map-time $t$, times rest-mass $m$, or $%
{\bf F}_{net}=\sum {\bf F}_i=m\frac{d{\bf w}}{dt}$. This net frame-variant
force is integrable for constant ``felt force'' ${\bf F}_o$, and simply
equal to $\frac{\Delta p}{\Delta t}=F_o$.

{\bf VIu. The unidirectional work-energy relation: }The rate at which an
object's energy changes, per unit map-distance traveled in the direction of
the ``felt'' force, equals the magnitude of the ``felt'' force $F_o$, i.e. $%
\frac{\Delta E}{\Delta x}=F_o$.

{\bf VIIu. The uni-directional chase-plane: }If one considers time $T$ on
the clocks of a chase-plane following our object with a velocity $v^{\prime
}=c\sqrt{1-\frac{\left( v/c\right) ^2}{2(\gamma -1)}}$, then uniformly
accelerated motion {\em at any speed} is described by Galileo's two simple
integrals for motion under constant acceleration, namely: $\Delta V=\alpha
\Delta T$ and $\Delta \left( V^2\right) =2\alpha \Delta x$ \cite{Noncoord}.  
Since chase-plane motion is intermediate to that of accelerated-object
and map-frame, at speeds low enough to make elapsed-times for map and object
similar, these equations predict the behavior of their clocks and velocities
as well.

\section{Summary and Discussion}

This paper explores a ``map-based'' approach to describing motion at any
speed, which follows from kinematical and dynamical postulates tied to the
metric equation alone. It begins by defining a set of minimally-variant 
variables for examining motion in flat space-time, from the vantage point of 
a chosen inertial frame.  Synchrony-free proper-velocity plays a powerful 
role throughout.  Three time and one spatial integral of constant proper 
acceleration then follow, including a new integral which makes use of 
time-elapsed on the clocks of a "Galilean chase-plane", traveling so as to  
ensure that Galileo's original equations of constant acceleration 
are honored at high speed as well.  The consequences of these integral 
equations, along with frame-specific postulates which come from conservation 
of the energy-momentum 4-vector, are re-arranged in a way which reflects 
the symmetries (and lack thereof) in a map-based view of space-time, while 
building on natural patterns already present in the classical (low-speed) 
``map-based laws'' format of Newton.  A set of symmetric equations (original as 
far as we know) for converting between proper-force and frame-variant force 
are offered.  These are applied to the simple case of the radial frame-variant 
electromagnetic force between a fixed and a moving charge, yielding a  
simple expression for the velocity-dependent proper-force between 
those two charges.  This proper-force, of course, predicts both the electrostatic 
and magnetic interactions between the two charges, in a form sharable 
(by virtue of its frame-invariance) between frames of motion.
 
Concerning immediate application in physics education (the author's area of 
interest), the strategy described here allows concrete solution of high-speed problems 
in flat space-time by problem-solvers not equipped for Lorentz transforms or the study 
of multiple inertial frames. By operationally defining time {\em only} with
respect to a chosen frame of reference, we can thus introduce students to the
study of motion correctly the first time using tools that work at any speed
\cite{anticipation}. Even the (3+1)D rules discussed here lend themselves to
easy summation (Fig. 1).

Thus it is useful to apply the metric equation from the vantage point of a
single and convenient, but otherwise arbitrarily chosen, reference frame.
Another way to think of this is to note that, although there is no
universally preferred frame, choosing problems in the study of high-speed
motion quite naturally leads to preferred points of view. For example, the
preferred frame for measuring the length of an object is the frame in which
that object is at rest. Using the notation of this paper, we list in Table I
some other preferences that follow naturally, whenever a problem involving
motion at high speeds is defined. We have not considered rotational motion in
this context, but expect that a ``map-based'' approach adaptable to motion 
at arbitrary speeds may be useful there as well.  Perhaps the rate of change 
{\em per unit traveler-time} of map-based angle will assume an analogous role 
to that of proper-velocity here.

Because the ``map-based'' rules described here stem directly from the metric
equation, they may be useful in non-flat space-time applications. For
example, given that we can describe the electromagnetic force in
frame-invariant form, a direct comparison to the proper-force expression for
Schwarzchild-metric gravity should be instructive and understandable by
students. Moreover, telling them that we can get gravity on earth by a
part-per-billion error in the space-time version of Pythagoras' theorem
should pique their interest in figuring out how as well!  Results of 
the distinctions drawn here are hardly beyond everyday experience:  
We go through life on earth experiencing a downward {\em proper-acceleration}, 
applied to each part of our being, whose consequences must be cancelled 
by an upward {\em frame-variant} force, applied to the soles of our feet if we 
wish to stand and not fall.  

For curved-space situations in which moving masses are sufficiently 
small that we can consider metric coefficients to result only from masses 
fixed with respect to a system of locally-inertial map-frames, some 
possible ``test-mass'' trajectories might yield themselves to
closed-form solution. Such trajectory solutions, in the spirit of Taylor 
and Wheeler \cite{Taylor2}, could provide students with another 
quantitative calculus-based handle on experiences associated with curved 
space-time.

\acknowledgments

My thanks for input relevant to this paper from W. A. Shurcliff,
E. F. Taylor, and A. A. Ungar. The work has benefited indirectly from
support by the U.S. Department of Energy, the Missouri Research Board, as
well as Monsanto and MEMC Electronic Materials Companies. It has benefited
most, however, from the interest and support of students at UM-St. Louis.

\onecolumn

\begin{table}[tbp] \centering
\begin{tabular}{cc}
{\bf Least variant quantity} & {\bf More subjective quantity} \\ \hline\hline
{proper (rest-frame) length $L_o$} & {coordinate-length $L=\frac{L_o}\gamma $}
\\ 
{proper (traveler) time elapsed $\Delta \tau $} & {coordinate-time elapsed $%
\Delta t=\gamma \Delta \tau $} \\ 
{proper-velocity ${\bf w}\equiv \frac{d{\bf x}}{d\tau }$} & 
{coordinate-velocity ${\bf v}\equiv \frac{d{\bf x}}{dt}=\frac{{\bf w}}\gamma $}
\\ 
{proper-acceleration ${\bf \alpha }$} & {coordinate-acceleration ${\bf a}=\frac{%
\gamma _{\perp }}{\gamma ^3}{\bf \alpha }$} \\ 
{proper-force ${\bf F}_o\equiv m{\bf \alpha }$} & {frame-variant force ${\bf F}%
=F_o\left( \frac 1{\gamma _{\perp }}{\bf i}_{\parallel }+\gamma _{\perp }%
\frac{v_{\perp }}c\frac{v\parallel }c{\bf i}_{\perp }\right) $} \\ \hline
{none} & {dynamically-defined momentum $m{\bf w}$ } \\ 
{none} & {dynamically-defined energy $\gamma mc^2$ } \\ \hline
\end{tabular}
\caption{Some less, and more, frame-dependent quantities in the study of high speed motion.
\label{Table1}}
\end{table}

\begin{figure}
\epsfxsize=8.5cm
\epsfysize=8.5cm
\caption{A compact summary of some rules of (3+1)D motion at any speed in flat space-time.}
\label{Fig1}
\end{figure}

\end{document}